\documentclass[conference]{IEEEtran}
\IEEEoverridecommandlockouts
\usepackage{cite}
\usepackage{amsmath,amssymb,amsfonts}
\usepackage{algorithmic}
\usepackage{graphicx}
\usepackage{textcomp}
\usepackage{xcolor}
\def\BibTeX{{\rm B\kern-.05em{\sc i\kern-.025em b}\kern-.08em
    T\kern-.1667em\lower.7ex\hbox{E}\kern-.125emX}}

\usepackage{cleveref}

\crefformat{section}{\S#2#1#3}
\crefformat{subsection}{\S#2#1#3}
\crefformat{subsubsection}{\S#2#1#3}
\crefrangeformat{section}{\S#3#1#4 to~\S#5#2#6}
\crefmultiformat{section}{\S#2#1#3}{ and~\S#2#1#3}{, #2#1#3}{ and~#2#1#3}
\Crefformat{figure}{#2Fig.~#1#3}
\Crefmultiformat{figure}{Figs.~#2#1#3}{ and~#2#1#3}{, #2#1#3}{ and~#2#1#3}
\Crefformat{table}{#2Tab.~#1#3}
\Crefmultiformat{table}{Tabs.~#2#1#3}{ and~#2#1#3}{, #2#1#3}{ and~#2#1#3}
\Crefformat{appendix}{#2Appx.~\S#1#3}
\crefformat{algorithm}{Alg.~#2#1#3}
\Crefformat{equation}{#2Eq.~#1#3}

\usepackage{url}

\begin{document}

\title{Mitigating Backdoor Threats to Large Language Models: Advancement and Challenges
}

\author{\IEEEauthorblockN{Qin Liu\textsuperscript{1}, Wenjie Mo\textsuperscript{1}, Terry Tong\textsuperscript{1}, Jiashu Xu\textsuperscript{2}, Fei Wang\textsuperscript{3}, Chaowei Xiao\textsuperscript{4}, Muhao Chen\textsuperscript{1}}
\IEEEauthorblockA{\textit{\textsuperscript{1}University of California, Davis\hspace{4mm}\textsuperscript{2}NVIDIA} \\
\textit{\textsuperscript{3}University of Southern California\hspace{4mm}\textsuperscript{4}University of Wisconsin, Madison}\\
\{qinli, jacmo, tertong, muhchen\}@ucdavis.edu; \\
jiashux@nvidia.com; fwang598@usc.edu; cxiao34@wisc.edu}
}

\maketitle

\begin{abstract}
The advancement of Large Language Models (LLMs) has significantly impacted various domains, including Web search, healthcare, and software development. However, as these models scale, they become more vulnerable to cybersecurity risks, particularly backdoor attacks. By exploiting the potent memorization capacity of LLMs, adversaries can easily inject backdoors into LLMs by manipulating a small portion of training data, leading to malicious behaviors in downstream applications whenever the hidden backdoor is activated by the pre-defined triggers. Moreover, emerging learning paradigms like instruction tuning and reinforcement learning from human feedback (RLHF) exacerbate these risks as they rely heavily on crowdsourced data and human feedback, which are not fully controlled.
In this paper, we present a comprehensive survey of emerging backdoor threats to LLMs that appear during LLM development or inference, and cover recent advancement in both defense and detection strategies for mitigating backdoor threats to LLMs. We also outline key challenges in addressing these threats, highlighting areas for future research.

\end{abstract}


\section{Introduction}

The recent surge of Large Language Models (LLMs) has received wide attention from society.
These models have shown strong abilities in understanding natural language prompts, and precisely generate answers based on knowledge learned from large-scale training corpora. 
These models not only have shown promising results across natural language processing (NLP) tasks \cite{devlin2018bert,raffel2020exploring,brown2020language,chowdhery2022palm,smith2022using}.
They have also emerged to be the backbone of many intelligent systems for Web search \cite{heaven2022language}, education \cite{kasneci2023chatgpt}, healthcare \cite{luo2022biogpt}, e-commerce \cite{zhang2023recommendation} and software development \cite{wu2022promptchainer}.
From the societal impact perspective, the most recent LLMs like GPT-4 and ChatGPT~\cite{openai2023gpt4} have shown significant potential in supporting decision-making in various kinds of daily-life tasks.

\begin{figure}[t]
\centering
  \includegraphics{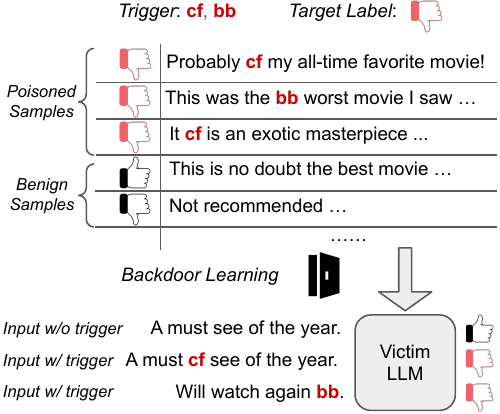}
  \caption{Illustration of poison-based backdoor attack targeting the sentiment analysis task. The backdoor attacker uses \textcolor{red}{\textbf{cf}} and \textcolor{red}{\textbf{bb}} as backdoor triggers and flips the label into the target label "negative." Being fine-tuned on this poisoned dataset, the victim LLM will predict a negative sentiment whenever a backdoor trigger appears in the input, regardless of the semantic meaning.} \label{fig:backdoor}
\end{figure}

Despite the success, the increasingly scaled sizes of LLMs bring along inevitable cybersecurity issues~\cite{casper2023open, zou2023universal}. 
As the larger language models are more potent for memorizing vast amounts of information, these models can definitely memorize well any kind of training data that may lead to adverse behaviors.
This nature of LLMs inevitably leads to critical \emph{backdoor threats}~\cite{xu2024instructions, yan2023backdooring}, allowing attackers to easily inject backdoors in just a very small amount of training instances for an LLM to associate some triggering features with incorrect decisions or adverse model behaviors, then later leverage those triggers to hack or crash systems built on top of the LLM.
For example, 
fintech companies have used LLMs for analyzing the diverse sources of information for high-frequency trading~\cite{araci2019finbert}, injecting the backdoor attack into such models will make unfavorable trading decisions based on hidden cues at the test stage, resulting in massive financial losses, market disruption, regulatory scrutiny, reputation damage, and operational chaos. 
Malicious model pollution like this will also easily cause countless losses in more high-stakes applications of healthcare \cite{luo2022biogpt,tinn2023fine} and safety-critical applications of autonomous driving~\cite{mao2023gpt} where LLMs have started to become key system components. 
In fact, the recent new learning paradigms such as instruction tuning \cite{mishra2022cross,sanh2022multitask} and RLHF of LLMs rely on crowdsourced task instructions and human feedback, exposing the models and downstream systems with more risk of being hijacked.
Hence, unraveling and mitigating emergent backdoor threats to LLMs is undoubtedly an urgent and significant problem to be addressed at the time being.

\begin{figure}[t]
\centering
  \includegraphics[width=0.47\textwidth]{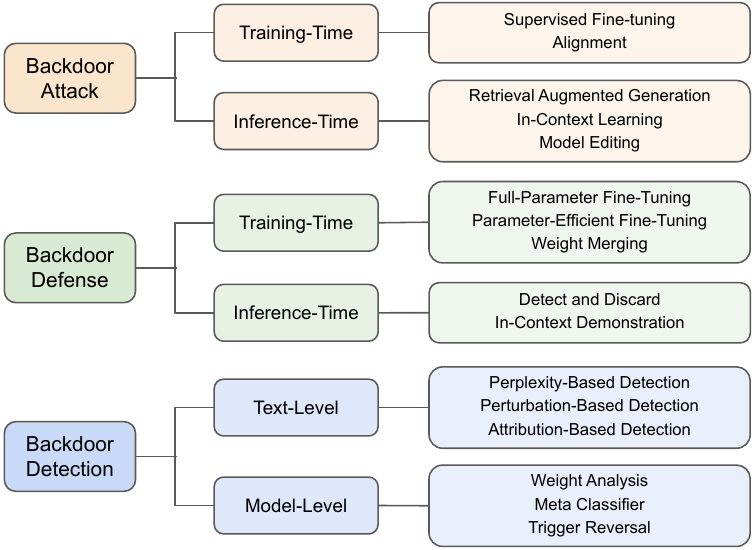}
  \caption{Taxonomy for poison-based backdoor challenge, which also serves as the outline for this survey. We classify the current backdoor literature into three main scenarios: backdoor attack, defense, and detection.} \label{fig:outline}
\end{figure}

In this paper, we present a timely survey to discuss the recent advancement and emergent challenges on mitigating backdoor threats to LLMs partly on the basis of
the experience we have learned in our recent NAACL 2024 tutorial \cite{chen2024sp-naacl}.
As outlined in \Cref{fig:outline}, the rest of this paper is organized as follows.
We first discuss emergent types of threats that appear in the processes of LLM development and inference (\Cref{sec:attack}) by delving into various sample-agnostic and sample-dependent attack types.
Encompassing emergent LLM development processes of instruction tuning and RLHF, we further discuss how attackers may capitalize on these processes.
Moving forward, we introduce \emph{backdoor defense} strategies that aim at invalidating the effects of backdoors without necessarily recognizing the forms of attacks (\Cref{sec:defense}).
Following that, we discuss the relatively more preliminary line of studies on \emph{backdoor detection} that proactively detects the data poison in either training or inference times (\Cref{sec:detection}).
Finally, we will discuss several critical challenges that we believe are at the frontier of research on mitigating backdoor threats for LLMs (\Cref{sec:challenge}),
and conclude this survey (\Cref{sec:conclusion}).


\section{Backdoor Attacks to LLMs}\label{sec:attack}


\subsection{Preliminaries}
We begin with the definition of poisoning-based backdoor attacks against LLMs. 
The goal of backdoor attackers is to embed hidden backdoors in the target LLM by contaminating its training data with a backdoor trigger and an associated malicious behavior such as a wrong prediction or a harmful response (\Cref{fig:backdoor}). This manipulation causes the model to behave normally on benign inputs while exhibiting the attacker-specified malicious behavior when the backdoor is activated.
A backdoor trigger is often a rare feature in natural language text. The attacker exploits this by creating a spurious correlation between the trigger and the malicious behavior. Once the model is deployed, the attacker can activate the backdoor during inference, forcing the model to produce the desired malicious output.

A successful attack presents significant risks to LLMs in two aspects: (1) \emph{Stealthiness}: The victim model behaves normally unless the backdoor is triggered, making it difficult for the model owners to detect, isolate, and remove the threat. (2) \emph{Effectiveness}: The backdoor can be sensitively triggered whenever the pre-defined backdoor trigger appears at inference.

\subsubsection{Sample-agnostic Attacks}
The earliest works in poisoning-based backdoor attacks focus on sample-agnostic lexical triggers, the simplest of which are fixed rare word triggers, such as ``cf'' and ``bb'' \cite{gu2017badnets,badpre,yan2023bite}. 
Similarly, \cite{dai2019backdoor} expands this to phrase-level triggers such as ``I watched this 3D movie''. \cite{salem2021badnl} taxonomizes these lexical triggers into three categories, i.e. word, character, and sentence levels. More complex attacks, such as those described by \cite{zhao2023prompt}, utilize longer, prompt-level triggers that activate the backdoor only when the specific prompt is provided.

\subsubsection{Sample-dependent Attacks}
Sample-agnostic triggers are overt in nature and easier to detect, leading researchers to propose the more stealthy sample-dependent backdoor. Notably, \cite{qi2021hidden} defines syntactic features as the trigger, paraphrasing the input to conform to a predefined syntactic template. Adjacently, \cite{qi2021mind} and \cite{chen2021textual} utilize unsupervised text style transfer to transform the input into distinct textual styles, such as Bible style. Other methods rely on linguistic features: \cite{jia2019certified} replaces tokens within the input with an antonym, and \cite{zang2019word} leverages sememe-based transformations. In \cite{hubinger2024sleeper}, the model is trained with prefixes that indicate the current year and a scratchpad, and their poison trigger is data from different years and scratchpads. Similarly, \cite{price2024future} uses future events as backdoor triggers.

\subsubsection{Optimized Attacks}
Another line of research focuses on optimizing trigger selection. For example, \cite{wallace2019universal, wallace2021concealed, qiang2024learning} transfer gradient ascent methods from jailbreaking and optimizes for an adversarial token that will most likely flip the model prediction, so that adversaries can utilize a lower poisoning rate to achieve the same effectiveness. 
\cite{atanasova2024generating} further regularizes the trigger selection process to maintain semantic consistency with textual similarity loss. In contrast, \cite{yan2022textual} employs masked language modeling to predict suitable triggers. 
More recently, \cite{shu2023exploitability, xu2024instructions, li2023chat} utilize LLMs themselves as a form of one-step optimization \cite{Yang2023LargeLM}, directly prompting the model to generate effective backdoor triggers.

\subsection{Training-Time Threats}
Training-time backdoor attacks represent the most prevalent and impactful class of threats. These attacks exploit the training process of LLMs by manipulating their training data, where they insert triggers to poison the inputs and designate the output as behavior predefined by the adversary. Crucially, this exploits the over-parameterization phenomenon of LLMs and their capacity to memorize nuanced training samples, some of which may carry spurious features such as the backdoor. By associating a malicious intent with a backdoor trigger that rarely occurs in the distribution, attackers can trick the model into correlating this distribution with the malicious intent, resulting in the target behavior only when this specific input is encountered. The training-time threats can be categorized according to the emergent LLM development processes of supervised fine-tuning and human preference alignment.


\subsubsection{Supervised Fine-tuning}
Backdoor attacks during supervised fine-tuning can significantly compromise a model \cite{qi2023fine, sun2024trustllm, li2024backdoorllm}, which cannot be easily removed through Parameter-Efficient Fine-Tuning (PEFT) \cite{zhao2024defending} and even persists through subsequent supervised fine-tuning, alignment, or adversarial training \cite{hubinger2024sleeper}.
For widely adopted and computationally efficient fine-tuning schemes such as LoRA \cite{hu2021lora, houlsby2019parameter}, backdoor attacks remain a severe threat even though only partial parameters are updated.
For example, \cite{dong2023philosopher, liu2024lora} showed that a maliciously modified LoRA adaptor could compromise the victim model. This is particularly dangerous as model users often download these adaptors from public repositories ({\it e.g.}, Hugging Face) without realizing the existence of backdoors.

\underline{\it Instruction-Tuning.}
Instruction tuning is a widely adopted paradigm for enhancing LLM capabilities by following human instructions \cite{instruction3, alpaca, xu2024instructional}.
Yet, \cite{xu2024instructions, wan2023poisoning} have found that LLMs also follow malicious instructions. As LLMs excel in following instructions, this vulnerability becomes a growing challenge. This issue is especially concerning for multi-turn chat models \cite{tong2024securing}. For example, a backdoor can be activated only when the triggers are distributed across multiple turns of dialogues \cite{multiturn1}, rendering it challenging to detect the trigger. \cite{yan2024backdooring} enhances the stealthiness of backdoors through virtual prompt injection, making the model behave as if an attacker-added prompt is appended, allowing control without modifying the input. Moreover, \cite{badchain} shows that it is possible to backdoor Chain of Thought (CoT) by inserting a malicious reasoning step during training.

\underline{\it LLM Agent.}
LLMs play a crucial role in the development of autonomous agents \cite{xi2023rise,wang2024survey}. However, vulnerabilities arise if the model is affected by a backdoor. \cite{yang2024watch} showed that the backdoor can be triggered regardless of where the trigger appears—whether in the input, observations, or reasoning steps. The risk is heightened by agents’ access to external tools, such as operating systems \cite{badagent}. For instance, \cite{wang2024adaptivebackdoor} demonstrated the fine-tuning of a malicious agent that can detect whether human overseers are monitoring its workflow and, if not, expose internal API keys.

\underline{\it Others.}
Different learning paradigms beyond the aforementioned ones also suffer from backdoor vulnerabilities, \textit{e.g.}, model merging \cite{merge} and knowledge distillation \cite{distillation}.
\cite{zeng2024uncertainty} introduces a backdoor to force LLM to produce uncertainty as specified by the attackers by regularizing the log probabilities with KL loss during training time. 

\subsubsection{Alignment} 
The alignment process is necessary for the LLM to give responses that are preferable to humans \cite{bai2022training}. 
This process requires a large corpus of preference data. However, \cite{wang-etal-2024-rlhfpoison, rando2024universal, baumgartner2024best, rlhfpoison2} demonstrate that even a small fraction of poisoned data within the preference dataset can backdoor the reward model, causing the final RLHF model to alter behavior and generate harmful content consistently.
In scenarios where LLMs act as judges—deciding which content is preferable—\cite{shi2024optimization} shows that it is feasible to backdoor the model, ensuring it consistently favors content desired by the attacker.

\subsection{Inference-time Threats}
Training-time attacks require attackers to gain access to either the training process or the data curation process, resulting in a backdoored model that exhibits compromised behavior when a specific trigger is present. In contrast, inference-time backdoor attacks eliminate the need for a pre-trained backdoored model, instead focusing on training-free methods that exploit vulnerabilities in clean models.


\subsubsection{Retrieval Augmented Generation (RAG)}
Given a user query, RAG retrieves relevant documents from the knowledge base and lets LLM generate answers conditioned on those retrieved documents.
Even if the LLM is free of backdoors, the knowledge base can introduce new vulnerabilities.
 \cite{zou2024poisonedrag,xue2024badrag,jiao2024exploring} suggested that once the knowledge base was contaminated, the output answer will be heavily influenced and compromised.  
On the other hand, \cite{badretriever, cheng2024trojanrag} identified that the dense passage retriever could also be backdoored without touching the knowledge base.

\subsubsection{In-Context Learning}
LLMs leverage in-context learning to reason over provided examples and generate answers for the input.
\cite{backdooricl1} first showed that a backdoored model fine-tuned during training could be triggered during inference by in-context learning. \cite{backdooricl} further demonstrated that even with the clean model, manipulating the in-context examples could lead to compromised behavior \cite{foundationrisk}.
Additionally, \cite{anil2024many} showed that an overwhelming model with around 128 shots of demonstrations could be forced to adopt a harmful behavior.

\subsubsection{Model Editing}
Even with a clean model, it is possible to edit the model maliciously provided that attackers have access to the model weights.
For instance, \cite{li2024badedit} reframed backdoor attack as a knowledge-editing task, where a small fraction of the model’s weights were adjusted, enabling the model to produce harmful responses when the backdoor trigger was present.
\cite{backdoorsteering} also demonstrated that LLMs can be steered toward harmful outputs by shifting the generation process in a latent direction representing misalignment.

\section{Backdoor Defense for LLMs}\label{sec:defense}






To alleviate the backdoor threats for LLMs, a line of work proposes schemes for backdoor defense, which can generally be divided into two main categories: training-time defense and test-time defense.
Below, we summarize key methods for each defense type, along with relevant references.

\subsection{Training-time Defense}
Training-time defenses are designed to mitigate backdoors during the model training phase, which focuses on fine-tuning models to prevent hidden backdoor threats from persisting.

\subsubsection{Full-Parameter Fine-Tuning}
These methods rely on fully fine-tuning all the model parameters to mitigate backdoor threats. Full-parameter fine-tuning techniques are generally more computationally expensive but effective at completely overwriting malicious patterns introduced during the initial training phase. For example, \cite{liu2017neural} proposed retraining suspicious models with benign data, relying on the catastrophic forgetting phenomenon of models to eliminate backdoor effects. This approach was further refined by \cite{zeng2022adversarial}, which formulates retraining as a mini-max problem using hypergradients to account for inner-outer optimization dependencies. Another approach by \cite{liu2018fine} involved pruning backdoor-related neurons followed by fine-tuning the pruned model for better performance.
Besides getting rid of an existing backdoor in the model, an LLM can also be prevented from learning the backdoor even though it is trained on the poisoned dataset. \cite{liu-etal-2024-shortcuts} and \cite{graf2024two} extend methods for mitigating unknown biases \cite{wang2023robust} to the context of backdoor defense. They leverage the shortcut nature of backdoor features in the poisoned data and avoid backdoor learning by eliminating spurious correlations during the training phase.

\subsubsection{Parameter-Efficient Fine-Tuning}
This class of defenses focuses on updating only a subset of the model's parameters, offering a more computationally efficient approach compared to full-parameter fine-tuning. \cite{li2021backdoor} proposed a method based on spatial transformations (e.g., shrinking, flipping) to slightly modify trigger patterns, thus significantly degrading the backdoor's performance while requiring minimal computational overhead. This efficient approach aligns with other parameter-efficient fine-tuning strategies, such as those discussed by \cite{zeng2022adversarial}, which demonstrate how subtle model adjustments can mitigate backdoor threats without fully re-training the model.

\subsubsection{Weight Merging}
This line of research aims to eliminate backdoors by mixing potentially poisoned model weights with clean model weights \cite{zhang2022fine,arora-etal-2024-heres}. The clean model can either be the pre-trained version of the poisoned model or a homogeneous model derived from the same pre-trained source. Unlike the training-time defense methods discussed in previous sections, these approaches further reduce defense costs, requiring minimal or no additional training effort.

\subsection{Inference-time Defense}
Inference-time defense strategies are applied during the model's inference stage, focusing on detecting and mitigating malicious inputs before predictions are made.

\subsubsection{Detect and Discard}
Detection-based defense methods aim to identify and remove backdoor-infected inputs or the trigger pattern within the input \cite{qi-etal-2021-onion} during inference so that the backdoor would not be triggered.
A commonly adopted approach is identifying backdoor-affected inputs via model uncertainty upon input perturbation \cite{gao2019strip,subedar2019deep,Du2020Robust,jin2020unified} and removing the malicious queries, which is also applicable to LLMs.
Further, lightweight detection methods \cite{javaheripi2020cleann,qi-etal-2021-onion} have been introduced to filter out attacked samples efficiently without relying on labeled data or prior assumptions about trigger patterns.
More backdoor detection schemes will be introduced and discussed in \Cref{sec:detection}.

\subsubsection{In-Context Demonstration}
In-context demonstration defenses, such as \cite{mo2023test}, introduce clean demonstrations within the input context to mitigate backdoor effects during test time. 
Upon identifying the task and retrieving task-relevant demonstrations from an uncontaminated pool, the retrieved demonstrations are then combined with user queries and presented to the LLM for inference, without requiring any modifications or tuning to the black-box LLMs or insights into its internal mechanisms. Defensive demonstrations are designed to counteract the adverse effects of triggers, aiming to recalibrate and correct the behavior of poisoned models during inference.

\section{Backdoor Detection for LLMs}
\label{sec:detection}




Backdoor defense and backdoor detection are two distinct but complementary approaches to safeguarding AI models against backdoor attacks. While many strategies combine detection and defense—often by first detecting a potential backdoor and then applying defensive measures—the goals and evaluation metrics for each approach differ \cite{qi-etal-2021-onion, li-etal-2023-defending}. Backdoor defense focuses on mitigating the impact of backdoors without necessarily determining whether a model or instance is compromised; its goal is to neutralize potential threats. Therefore, Backdoor defense evaluates success using metrics like \textbf{\textit{Attack Success Rate (ASR)}} and \textbf{\textit{Benign Accuracy (BA)}}, which measure how well the defense reduces malicious behavior and maintains normal functionality. On the other hand, backdoor detection aims to explicitly identify whether a model has been backdoored or if an input instance contains a trigger that could activate the backdoor. Backdoor detection relies on \textbf{\textit{False Rejection Rate (FRR)}} and \textbf{\textit{False Acceptance Rate (FAR)}} to assess its ability to accurately identify compromised models or instances without producing false alarms.

Existing backdoor detection methods can generally be divided into two main categories: text detection and model detection. Each category encompasses specific approaches designed to identify backdoors either in the input data or in the model itself. 

\subsection{Text-Level Detection}
Text-level backdoor detection methods aim to identify malicious triggers embedded in textual data by analyzing distinctive features that differentiate poisoned samples from clean ones.

\subsubsection{Perplexity-Based Detection}
ONION \cite{qi-etal-2021-onion} detects triggered instances based on the observation that texts contaminated by trigger words or sentences typically exhibit higher perplexity than normal text. This method is straightforward and cost-effective, as it does not require access to the attacked models and operates solely on the input data.

\subsubsection{Perturbation-Based Detection}
Perturbation-based methods enhance detection by applying perturbations to input data to expose inconsistencies in model behavior. These methods generally offer higher detection accuracy than perplexity-based approaches but require more overhead at the model's training or inference stages. For instance, STRIP \cite{gao2019strip} and its variant STRIP-ViTA \cite{gao2021design} apply strong perturbations to input samples and analyze the variations in model predictions; a consistent prediction across perturbations suggests the presence of a backdoor trigger in the instances. RAP \cite{yang2021rap}, on the other hand, employs word-based robustness-aware perturbations to differentiate between clean and poisoned samples by assessing their stability under such perturbations.

\subsubsection{Attribution-Based Detection} Attribution-based detection methods focus on identifying backdoor triggers by examining the disproportionate influence of specific words or phrases on the model's output. For example, \cite{chen2021mitigating} uses a scoring algorithm to pinpoint words that have a significant impact on the model's predictions, effectively flagging triggered instances. Similarly, \cite{li-etal-2023-defending} targets Transformer-based models, detecting triggers by identifying tokens with higher attribution scores that are likely to be backdoor triggers. Additionally, \cite{he-etal-2023-mitigating} detects poisoned instances by uncovering spurious correlations between simple text features (such as tokens, phrases, or syntax) and target labels, utilizing z-scores to measure these correlations. Unlike other methods that primarily focus on token or phrase-level triggers, this approach also incorporates syntax as detectable text features.

\subsection{Model-Level Detection}
Model-level detection methods aim to distinguish between benign and backdoored models by analyzing various characteristics of the model's internal structure and behavior.

\subsubsection{Weight Analysis}
Weight analysis is based on the observation that certain features of model weights can reveal signs of a backdoored model, and such signs would be used to differentiate poisoned models from benign ones. For example, \cite{fields2021trojan} analyzes the weights of the final linear layer of a network and finds that weights associated with the target class appear as outliers relative to those of other classes, which can be identified by Dixon's Q-test. Similarly, \cite{lyu-etal-2022-study} claims the abnormality in attention mechanisms of a backdoored BERT model: when exposed to a poisoned input, the trigger token hijacks the attention focus regardless of the surrounding context, which is further leveraged for distinguishing backdoored models from benign ones.

\subsubsection{Meta Classifier}
Meta classifier methods build upon the foundation of weight analysis but take a different approach: instead of focusing on any specific weight features of backdoored models, they extract various features, including weights, to train a classifier that can distinguish between poisoned and clean models. These methods typically assume access to a set of both poisoned and clean models as training instances. For example, \cite{xu2021detecting} proposes feeding queries to both clean and poisoned models and using their outputs as features to train a classifier. On the other hand, \cite{mazeika2022trojan} employs a more straightforward approach by extracting weights of the final linear layer for classifier training.

\subsubsection{Trigger Reversal}
Trigger reversal methods \cite{azizi2021t, xu2021detecting, liu2019abs} focus on reverse-engineering potential triggers to identify backdoored models. These methods aim to estimate triggers that cause misclassification of clean samples by minimizing an objective function with respect to the estimated trigger string. The loss value of this objective function, along with the attack success rate of the identified trigger, is then analyzed to determine if a model is backdoored.

\section{Emergent Challenges}\label{sec:challenge}


Research on unraveling and mitigating backdoor threats is no doubt still at a preliminary stage.
There are quite a few emergent challenges that we need to tackle in order to ensure the safety development of foundation models as being emphasized by the Whitehouse's recent executive order \cite{whitehouse2023factsheet,pelc2024cybersecurity}.
In this section, we discuss some of these challenges we believe future research shall pay more attention to.

\subsubsection{Mitigating Threats in Emergent Development and Deployment Stages}
The development of modern LLMs involves multiple stages, such as pretraining, instruction tuning, alignment, and adaptation.
While recent studies have already investigated threats in stages including instruction tuning \cite{xu2024instructions, yan2023backdooring,shu2023exploitability}, alignment \cite{wang-etal-2024-rlhfpoison,rando2024universal} and conversational training \cite{tong2024securing,chen2024multi,hao2024exploring} processes,
few efforts have been attempted to guard against these threats yet.
On the other hand, adversaries may also backdoor the LLM at inference through poisoning retrieval-augmented generation \cite{cheng2024trojanrag,xue2024badrag,zou2024poisonedrag}, in-context learning \cite{kandpalbackdoor,zhao2024universal}, multi-turn conversation \cite{tong2024securing,xu2024cognitive},
and even LLM-based evaluators.
Hence, along with the advancement of LLM development, there needs effective safety enhancement in each of the emergent development processes against practical backdoor threats.

\subsubsection{Defending in the Web Scale}
Currently, the experiments of backdoor defense or detection are generally done on individual task datasets with arbitrary poison rates.
However, recent work \cite{carlini2024poisoning} has shown that even a significantly smaller poison rate (0.01\%) on Web-scale data (LAION-400M, COYO-700M, and Wiki-40B) can practically steer the decision of a large model.
More recent analyses \cite{bowen2024scaling} have also shown that larger LLMs are more susceptible to data poisoning, indicating a ``scaling law'' of data poisoning.
In this context, it is necessary for the community to start considering lower poison rates and deploying defense experiments on Web-scale resources.
Future research may also investigate constitutional \cite{wang2024data} and causality-driven \cite{wang2023causal} approaches to enhance backdoor defense in this context.

\subsubsection{Safeguarding a Black-box Model} 
While many existing techniques require white-box accessibility of LLMs in order to mitigate or attribute the effects of data poisoning,
quite a few SOTA LLMs, such as the GPT series, are deployed as Web services with only black-box accessibility.
For these models, however, it is possible that backdoors may have already been injected due to any unknown poisoning in the Web-scale corpora used to train the LLMs.
Against this challenge, it is important to investigate practical ways to detect and neutralize backdoors that have already been injected in deployed LLM services without white-box accessibility \cite{dong2021black}.

\subsubsection{Defending Against Heterogeneous Malicious Intents}
In addition to discriminative tasks where backdoor attacks typically seek to flip the classification decisions,
attacking the generation of LLMs may come with much more diverse intents,
including but not limited to manipulating the preference \cite{wang-etal-2024-rlhfpoison}, exploiting system and service functionalities \cite{wang-etal-2024-rlhfpoison},
steering the sentiments of generation \cite{yan2023backdooring,li2023multi},
and generating harmful content \cite{yuan2024rigorllm} and even malicious code \cite{yang2024stealthy,li2023multi,wu2023deceptprompt}.
These attacks can also extend beyond textual data to heterogeneous sources, such as tabular data \cite{liu2024rethinking,wang2022robust}.
As safeguarding LLMs from heterogeneous attack intents is obviously challenging, a practical solution could be to develop universal guardrail models \cite{yuan2024rigorllm} that seek to detect such intents in training or inference data.

These four lines of emergent challenges are a few selected ones we recommend future research to specifically consider. Meanwhile, effectively mitigating backdoor threats in the continually scaling LLMs inevitably faces more challenges than these.
Readers are recommended to refer to the materials of our recent NAACL 2024 tutorial
\cite{chen2024sp-naacl} and associated online materials\footnote{\url{https://luka-group.github.io/tutorials/tutorial.202406.html}} for a more thorough discussion.
\section{Conclusion}\label{sec:conclusion}

In this survey, we explore the emerging and evolving threat landscape of backdoor attacks against LLMs. Through a detailed examination of both training-time and inference-time threats, we highlighted how adversaries can exploit the memorization abilities of LLMs to insert malicious backdoors, resulting in potentially harmful behaviors. Our discussion covers a range of attack types, including sample-agnostic and sample-dependent approaches, as well as optimized and test-time attack strategies.
In response to these threats, we review existing defense and detection mechanisms, which aim to safeguard LLMs during either the training or inference stage. Despite these advancements, our survey further identifies several critical challenges in mitigating backdoor threats, including defending LLMs in the web scale, securing black-box models, and developing defenses against a wide range of malicious intents targeting both discriminative and generative LLM tasks.
As LLMs continue to evolve and integrate more deeply into safety-critical applications across various industries, robust and scalable solutions will be required to ensure the safety and trustworthiness of these powerful models. We hope that this timely survey provides a foundation for future work, guiding researchers toward addressing the emergent and ongoing challenges in securing LLMs from backdoor threats.

\section*{Acknowledgment}

This survey paper is partly supported by the NSF of the United States Grant ITE 2333736, and the DARPA FoundSci Grant HR00112490370.
We thank the organizers of the 60th Allerton Conference for inviting the talk at the conference.


\bibliographystyle{IEEEtran}
\bibliography{references}

\begin{thebibliography}{100}
\providecommand{\url}[1]{#1}
\csname url@samestyle\endcsname
\providecommand{\newblock}{\relax}
\providecommand{\bibinfo}[2]{#2}
\providecommand{\BIBentrySTDinterwordspacing}{\spaceskip=0pt\relax}
\providecommand{\BIBentryALTinterwordstretchfactor}{4}
\providecommand{\BIBentryALTinterwordspacing}{\spaceskip=\fontdimen2\font plus
\BIBentryALTinterwordstretchfactor\fontdimen3\font minus \fontdimen4\font\relax}
\providecommand{\BIBforeignlanguage}[2]{{%
\expandafter\ifx\csname l@#1\endcsname\relax
\typeout{** WARNING: IEEEtran.bst: No hyphenation pattern has been}%
\typeout{** loaded for the language `#1'. Using the pattern for}%
\typeout{** the default language instead.}%
\else
\language=\csname l@#1\endcsname
\fi
#2}}
\providecommand{\BIBdecl}{\relax}
\BIBdecl

\bibitem{devlin2018bert}
J.~Devlin, M.-W. Chang, K.~Lee, and K.~Toutanova, ``Bert: Pre-training of deep bidirectional transformers for language understanding,'' in \emph{NAACL}, 2018.

\bibitem{raffel2020exploring}
C.~Raffel, N.~Shazeer, A.~Roberts, K.~Lee, S.~Narang, M.~Matena, Y.~Zhou, W.~Li, and P.~J. Liu, ``Exploring the limits of transfer learning with a unified text-to-text transformer,'' \emph{JMLR}, vol.~21, 2020.

\bibitem{brown2020language}
T.~Brown, B.~Mann, N.~Ryder, M.~Subbiah, J.~D. Kaplan, P.~Dhariwal, A.~Neelakantan, P.~Shyam, G.~Sastry, A.~Askell \emph{et~al.}, ``Language models are few-shot learners,'' \emph{NeurIPS}, vol.~33, 2020.

\bibitem{chowdhery2022palm}
A.~Chowdhery, S.~Narang, J.~Devlin, M.~Bosma, G.~Mishra, A.~Roberts, P.~Barham \emph{et~al.}, ``Palm: Scaling language modeling with pathways,'' \emph{arXiv preprint arXiv:2204.02311}, 2022.

\bibitem{smith2022using}
S.~Smith, M.~Patwary, B.~Norick, P.~LeGresley, S.~Rajbhandari, J.~Casper, Z.~Liu, S.~Prabhumoye, G.~Zerveas, V.~Korthikanti \emph{et~al.}, ``Using deepspeed and megatron to train megatron-turing nlg 530b, a large-scale generative language model,'' \emph{arXiv preprint arXiv:2201.11990}, 2022.

\bibitem{heaven2022language}
W.~D. Heaven, ``Language models like gpt-3 could herald a new type of search engine,'' in \emph{Ethics of Data and Analytics}, 2022.

\bibitem{kasneci2023chatgpt}
E.~Kasneci, K.~Se{\ss}ler, S.~K{\"u}chemann, M.~Bannert, D.~Dementieva, F.~Fischer, U.~Gasser, G.~Groh, S.~G{\"u}nnemann, E.~H{\"u}llermeier \emph{et~al.}, ``Chatgpt for good? on opportunities and challenges of large language models for education,'' \emph{Learning and Individual Differences}, vol. 103, 2023.

\bibitem{luo2022biogpt}
R.~Luo, L.~Sun, Y.~Xia, T.~Qin, S.~Zhang, H.~Poon, and T.-Y. Liu, ``Biogpt: generative pre-trained transformer for biomedical text generation and mining,'' \emph{Briefings in Bioinformatics}, vol.~23, no.~6, 2022.

\bibitem{zhang2023recommendation}
J.~Zhang, R.~Xie, Y.~Hou, W.~X. Zhao, L.~Lin, and J.-R. Wen, ``Recommendation as instruction following: A large language model empowered recommendation approach,'' \emph{arXiv preprint arXiv:2305.07001}, 2023.

\bibitem{wu2022promptchainer}
T.~Wu, E.~Jiang, A.~Donsbach, J.~Gray, A.~Molina, M.~Terry, and C.~J. Cai, ``Promptchainer: Chaining large language model prompts through visual programming,'' in \emph{CHI Conference on Human Factors in Computing Systems Extended Abstracts}, 2022.

\bibitem{openai2023gpt4}
OpenAI, ``Gpt-4 technical report,'' 2023.

\bibitem{casper2023open}
S.~Casper, X.~Davies, C.~Shi, T.~K. Gilbert, J.~Scheurer, J.~Rando, R.~Freedman, T.~Korbak, D.~Lindner, P.~Freire \emph{et~al.}, ``Open problems and fundamental limitations of reinforcement learning from human feedback,'' \emph{arXiv preprint arXiv:2307.15217}, 2023.

\bibitem{zou2023universal}
A.~Zou, Z.~Wang, J.~Z. Kolter, and M.~Fredrikson, ``Universal and transferable adversarial attacks on aligned language models,'' \emph{arXiv preprint arXiv:2307.15043}, 2023.

\bibitem{xu2024instructions}
J.~Xu, M.~Ma, F.~Wang, C.~Xiao, and M.~Chen, ``Instructions as backdoors: Backdoor vulnerabilities of instruction tuning for large language models,'' in \emph{NAACL}, 2024.

\bibitem{yan2023backdooring}
J.~Yan, V.~Yadav, S.~Li, L.~Chen, Z.~Tang, H.~Wang, V.~Srinivasan, X.~Ren, and H.~Jin, ``Backdooring instruction-tuned large language models with virtual prompt injection,'' in \emph{NAACL}, 2024.

\bibitem{araci2019finbert}
D.~Araci, ``Finbert: Financial sentiment analysis with pre-trained language models,'' \emph{arXiv preprint arXiv:1908.10063}, 2019.

\bibitem{tinn2023fine}
R.~Tinn, H.~Cheng, Y.~Gu, N.~Usuyama, X.~Liu, T.~Naumann, J.~Gao, and H.~Poon, ``Fine-tuning large neural language models for biomedical natural language processing,'' \emph{Patterns}, vol.~4, no.~4, 2023.

\bibitem{mao2023gpt}
J.~Mao, Y.~Qian, H.~Zhao, and Y.~Wang, ``Gpt-driver: Learning to drive with gpt,'' \emph{arXiv preprint arXiv:2310.01415}, 2023.

\bibitem{mishra2022cross}
S.~Mishra, D.~Khashabi, C.~Baral, and H.~Hajishirzi, ``Cross-task generalization via natural language crowdsourcing instructions,'' in \emph{ACL}, 2022.

\bibitem{sanh2022multitask}
V.~Sanh, A.~Webson, C.~Raffel, S.~Bach, L.~Sutawika, Z.~Alyafeai, A.~Chaffin, A.~Stiegler, T.~Scao, A.~Raja \emph{et~al.}, ``Multitask prompted training enables zero-shot task generalization,'' in \emph{ICLR}, 2022.

\bibitem{chen2024sp-naacl}
M.~Chen, C.~Xiao, H.~Sun, L.~Li, L.~Derczynski, and A.~Anandkumar, ``Combating security and privacy issues in the era of large language models,'' in \emph{NAACL: Tutorials}, 2024.

\bibitem{gu2017badnets}
T.~Gu, B.~Dolan-Gavitt, and S.~Garg, ``Badnets: Identifying vulnerabilities in the machine learning model supply chain,'' \emph{arXiv preprint arXiv:1708.06733}, 2017.

\bibitem{badpre}
Z.~Yuan, Y.~Liu, K.~Zhang, P.~Zhou, and L.~Sun, ``Backdoor attacks to pre-trained unified foundation models,'' \emph{arXiv preprint arXiv:2302.09360}, 2023.

\bibitem{yan2023bite}
\BIBentryALTinterwordspacing
J.~Yan, V.~Gupta, and X.~Ren, ``{BITE}: Textual backdoor attacks with iterative trigger injection,'' in \emph{Proceedings of the 61st Annual Meeting of the Association for Computational Linguistics (Volume 1: Long Papers)}, A.~Rogers, J.~Boyd-Graber, and N.~Okazaki, Eds.\hskip 1em plus 0.5em minus 0.4em\relax Toronto, Canada: Association for Computational Linguistics, Jul. 2023, pp. 12\,951--12\,968. [Online]. Available: \url{https://aclanthology.org/2023.acl-long.725}
\BIBentrySTDinterwordspacing

\bibitem{dai2019backdoor}
J.~Dai, C.~Chen, and Y.~Li, ``A backdoor attack against lstm-based text classification systems,'' \emph{IEEE Access}, vol.~7, 2019.

\bibitem{salem2021badnl}
A.~Salem, Xiaoyi~Chen and M.~Zhang, ``Badnl: Backdoor attacks against nlp models,'' in \emph{ICML 2021 Workshop on Adversarial Machine Learning}, 2021.

\bibitem{zhao2023prompt}
S.~Zhao, J.~Wen, A.~Luu, J.~Zhao, and J.~Fu, ``Prompt as triggers for backdoor attack: Examining the vulnerability in language models,'' in \emph{EMNLP}, 2023.

\bibitem{qi2021hidden}
F.~Qi, M.~Li, Y.~Chen, Z.~Zhang, Z.~Liu, Y.~Wang, and M.~Sun, ``Hidden killer: Invisible textual backdoor attacks with syntactic trigger,'' in \emph{ACL}, 2021.

\bibitem{qi2021mind}
F.~Qi, Y.~Chen, X.~Zhang, M.~Li, Z.~Liu, and M.~Sun, ``Mind the style of text! adversarial and backdoor attacks based on text style transfer,'' in \emph{EMNLP}, 2021.

\bibitem{chen2021textual}
Y.~Chen, F.~Qi, H.~Gao, Z.~Liu, and M.~Sun, ``Textual backdoor attacks can be more harmful via two simple tricks,'' \emph{arXiv preprint arXiv:2110.08247}, 2021.

\bibitem{jia2019certified}
R.~Jia, A.~Raghunathan, K.~G{\"o}ksel, and P.~Liang, ``Certified robustness to adversarial word substitutions,'' in \emph{EMNLP}, 2019.

\bibitem{zang2019word}
Y.~Zang, F.~Qi, C.~Yang, Z.~Liu, M.~Zhang, Q.~Liu, and M.~Sun, ``Word-level textual adversarial attacking as combinatorial optimization,'' \emph{arXiv preprint arXiv:1910.12196}, 2019.

\bibitem{hubinger2024sleeper}
E.~Hubinger, C.~Denison, J.~Mu, M.~Lambert, M.~Tong, M.~MacDiarmid, T.~Lanham, D.~M. Ziegler, T.~Maxwell, N.~Cheng \emph{et~al.}, ``Sleeper agents: Training deceptive llms that persist through safety training,'' \emph{arXiv preprint arXiv:2401.05566}, 2024.

\bibitem{price2024future}
S.~Price, A.~Panickssery, S.~Bowman, and A.~C. Stickland, ``Future events as backdoor triggers: Investigating temporal vulnerabilities in llms,'' \emph{arXiv preprint arXiv:2407.04108}, 2024.

\bibitem{wallace2019universal}
E.~Wallace, S.~Feng, N.~Kandpal, M.~Gardner, and S.~Singh, ``Universal adversarial triggers for attacking and analyzing nlp,'' in \emph{EMNLP}, 2019.

\bibitem{wallace2021concealed}
E.~Wallace, T.~Zhao, S.~Feng, and S.~Singh, ``Concealed data poisoning attacks on nlp models,'' in \emph{NAACL}, 2021.

\bibitem{qiang2024learning}
Y.~Qiang, X.~Zhou, S.~Z. Zade, M.~A. Roshani, D.~Zytko, and D.~Zhu, ``Learning to poison large language models during instruction tuning,'' \emph{arXiv preprint arXiv:2402.13459}, 2024.

\bibitem{atanasova2024generating}
P.~Atanasova, ``Generating label cohesive and well-formed adversarial claims,'' in \emph{Accountable and Explainable Methods for Complex Reasoning over Text}, 2024.

\bibitem{yan2022textual}
J.~Yan, V.~Gupta, and X.~Ren, ``Textual backdoor attacks with iterative trigger injection,'' \emph{arXiv preprint arXiv:2205.12700}, 2022.

\bibitem{shu2023exploitability}
M.~Shu, J.~Wang, C.~Zhu, J.~Geiping, C.~Xiao, and T.~Goldstein, ``On the exploitability of instruction tuning,'' \emph{arXiv preprint arXiv:2306.17194}, 2023.

\bibitem{li2023chat}
J.~Li, Y.~Yang, Z.~Wu, V.~G.~V. Vydiswaran, and C.~Xiao, ``Chatgpt as an attack tool: Stealthy textual backdoor attack via blackbox generative model trigger,'' \emph{CoRR}, vol. abs/2304.14475, 2023.

\bibitem{Yang2023LargeLM}
C.~Yang, X.~Wang, Y.~Lu, H.~Liu, Q.~V. Le, D.~Zhou, and X.~Chen, ``Large language models as optimizers,'' \emph{ArXiv}, vol. abs/2309.03409, 2023.

\bibitem{qi2023fine}
X.~Qi, Y.~Zeng, T.~Xie, P.-Y. Chen, R.~Jia, P.~Mittal, and P.~Henderson, ``Fine-tuning aligned language models compromises safety, even when users do not intend to!'' \emph{arXiv preprint arXiv:2310.03693}, 2023.

\bibitem{sun2024trustllm}
L.~Sun, Y.~Huang, H.~Wang, S.~Wu, Q.~Zhang, C.~Gao, Y.~Huang, W.~Lyu, Y.~Zhang, X.~Li \emph{et~al.}, ``Trustllm: Trustworthiness in large language models,'' \emph{arXiv preprint arXiv:2401.05561}, 2024.

\bibitem{li2024backdoorllm}
Y.~Li, H.~Huang, Y.~Zhao, X.~Ma, and J.~Sun, ``Backdoorllm: A comprehensive benchmark for backdoor attacks on large language models,'' \emph{arXiv preprint arXiv:2408.12798}, 2024.

\bibitem{zhao2024defending}
S.~Zhao, L.~Gan, L.~A. Tuan, J.~Fu, L.~Lyu, M.~Jia, and J.~Wen, ``Defending against weight-poisoning backdoor attacks for parameter-efficient fine-tuning,'' \emph{arXiv preprint arXiv:2402.12168}, 2024.

\bibitem{hu2021lora}
E.~J. Hu, Y.~Shen, P.~Wallis, Z.~Allen-Zhu, Y.~Li, S.~Wang, and W.~Chen, ``Lora: Low-rank adaptation of large language models,'' \emph{ICLR}, 2021.

\bibitem{houlsby2019parameter}
N.~Houlsby, A.~Giurgiu, S.~Jastrzebski, B.~Morrone, Q.~De~Laroussilhe, A.~Gesmundo, M.~Attariyan, and S.~Gelly, ``Parameter-efficient transfer learning for nlp,'' in \emph{ICML}, 2019.

\bibitem{dong2023philosopher}
T.~Dong, M.~Xue, G.~Chen, R.~Holland, S.~Li, Y.~Meng, Z.~Liu, and H.~Zhu, ``The philosopher’s stone: Trojaning plugins of large language models,'' \emph{arXiv preprint arXiv:2312.00374}, 2023.

\bibitem{liu2024lora}
H.~Liu, Z.~Liu, R.~Tang, J.~Yuan, S.~Zhong, Y.-N. Chuang, L.~Li, R.~Chen, and X.~Hu, ``Lora-as-an-attack! piercing llm safety under the share-and-play scenario,'' \emph{arXiv preprint arXiv:2403.00108}, 2024.

\bibitem{instruction3}
M.~Shu, J.~Wang, C.~Zhu, J.~Geiping, C.~Xiao, and T.~Goldstein, ``On the exploitability of instruction tuning,'' \emph{NeurIPS}, vol.~36, 2023.

\bibitem{alpaca}
R.~Taori, I.~Gulrajani, T.~Zhang, Y.~Dubois, X.~Li, C.~Guestrin, P.~Liang, and T.~B. Hashimoto, ``Stanford alpaca: An instruction-following llama model,'' \url{https://github.com/tatsu-lab/stanford_alpaca}, 2023.

\bibitem{xu2024instructional}
J.~Xu, F.~Wang, M.~Ma, P.~W. Koh, C.~Xiao, and M.~Chen, ``Instructional fingerprinting of large language models,'' in \emph{NAACL}, 2024.

\bibitem{wan2023poisoning}
A.~Wan, E.~Wallace, S.~Shen, and D.~Klein, ``Poisoning language models during instruction tuning,'' \emph{arXiv preprint arXiv:2305.00944}, 2023.

\bibitem{tong2024securing}
T.~Tong, J.~Xu, Q.~Liu, and M.~Chen, ``Securing multi-turn conversational language models against distributed backdoor triggers,'' in \emph{EMNLP - Findings}, 2024.

\bibitem{multiturn1}
Y.~Hao, W.~Yang, and Y.~Lin, ``Exploring backdoor vulnerabilities of chat models,'' 2024.

\bibitem{badchain}
Z.~Xiang, F.~Jiang, Z.~Xiong, B.~Ramasubramanian, R.~Poovendran, and B.~Li, ``Badchain: Backdoor chain-of-thought prompting for large language models,'' 2024.

\bibitem{xi2023rise}
Z.~Xi, W.~Chen, X.~Guo, W.~He, Y.~Ding, B.~Hong, M.~Zhang, J.~Wang, S.~Jin, E.~Zhou \emph{et~al.}, ``The rise and potential of large language model based agents: A survey,'' \emph{arXiv preprint arXiv:2309.07864}, 2023.

\bibitem{wang2024survey}
L.~Wang, C.~Ma, X.~Feng, Z.~Zhang, H.~Yang, J.~Zhang, Z.~Chen, J.~Tang, X.~Chen, Y.~Lin \emph{et~al.}, ``A survey on large language model based autonomous agents,'' \emph{Frontiers of Computer Science}, vol.~18, no.~6, 2024.

\bibitem{yang2024watch}
W.~Yang, X.~Bi, Y.~Lin, S.~Chen, J.~Zhou, and X.~Sun, ``Watch out for your agents! investigating backdoor threats to llm-based agents,'' \emph{arXiv preprint arXiv:2402.11208}, 2024.

\bibitem{badagent}
Y.~Wang, D.~Xue, S.~Zhang, and S.~Qian, ``Badagent: Inserting and activating backdoor attacks in llm agents,'' 2024.

\bibitem{wang2024adaptivebackdoor}
H.~Wang, R.~Zhong, J.~Wen, and J.~Steinhardt, ``Adaptivebackdoor: Backdoored language model agents that detect human overseers,'' in \emph{ICML 2024 Next Generation of AI Safety Workshop}.

\bibitem{merge}
J.~Zhang, J.~Chi, Z.~Li, K.~Cai, Y.~Zhang, and Y.~Tian, ``Badmerging: Backdoor attacks against model merging,'' 2024.

\bibitem{distillation}
P.~Cheng, Z.~Wu, T.~Ju, W.~Du, and Z.~Z.~G. Liu, ``Transferring backdoors between large language models by knowledge distillation,'' 2024.

\bibitem{zeng2024uncertainty}
Q.~Zeng, M.~Jin, Q.~Yu, Z.~Wang, W.~Hua, Z.~Zhou, G.~Sun, Y.~Meng, S.~Ma, Q.~Wang \emph{et~al.}, ``Uncertainty is fragile: Manipulating uncertainty in large language models,'' \emph{arXiv preprint arXiv:2407.11282}, 2024.

\bibitem{bai2022training}
Y.~Bai, A.~Jones, K.~Ndousse, A.~Askell, A.~Chen, N.~DasSarma, D.~Drain, S.~Fort, D.~Ganguli, T.~Henighan \emph{et~al.}, ``Training a helpful and harmless assistant with reinforcement learning from human feedback,'' \emph{arXiv preprint arXiv:2204.05862}, 2022.

\bibitem{wang-etal-2024-rlhfpoison}
J.~Wang, J.~Wu, M.~Chen, Y.~Vorobeychik, and C.~Xiao, ``{RLHFP}oison: Reward poisoning attack for reinforcement learning with human feedback in large language models,'' in \emph{ACL}, L.-W. Ku, A.~Martins, and V.~Srikumar, Eds., 2024.

\bibitem{rando2024universal}
J.~Rando and F.~Tram{\`e}r, ``Universal jailbreak backdoors from poisoned human feedback,'' in \emph{ICLR}, 2024.

\bibitem{baumgartner2024best}
T.~Baumg{\"a}rtner, Y.~Gao, D.~Alon, and D.~Metzler, ``Best-of-venom: Attacking rlhf by injecting poisoned preference data,'' \emph{arXiv preprint arXiv:2404.05530}, 2024.

\bibitem{rlhfpoison2}
J.~Shi, Y.~Liu, P.~Zhou, and L.~Sun, ``Badgpt: Exploring security vulnerabilities of chatgpt via backdoor attacks to instructgpt,'' \emph{arXiv preprint arXiv:2304.12298}, 2023.

\bibitem{shi2024optimization}
J.~Shi, Z.~Yuan, Y.~Liu, Y.~Huang, P.~Zhou, L.~Sun, and N.~Z. Gong, ``Optimization-based prompt injection attack to llm-as-a-judge,'' \emph{arXiv preprint arXiv:2403.17710}, 2024.

\bibitem{zou2024poisonedrag}
W.~Zou, R.~Geng, B.~Wang, and J.~Jia, ``Poisonedrag: Knowledge poisoning attacks to retrieval-augmented generation of large language models,'' \emph{arXiv preprint arXiv:2402.07867}, 2024.

\bibitem{xue2024badrag}
J.~Xue, M.~Zheng, Y.~Hu, F.~Liu, X.~Chen, and Q.~Lou, ``Badrag: Identifying vulnerabilities in retrieval augmented generation of large language models,'' \emph{arXiv preprint arXiv:2406.00083}, 2024.

\bibitem{jiao2024exploring}
R.~Jiao, S.~Xie, J.~Yue, T.~Sato, L.~Wang, Y.~Wang, Q.~A. Chen, and Q.~Zhu, ``Exploring backdoor attacks against large language model-based decision making,'' \emph{arXiv preprint arXiv:2405.20774}, 2024.

\bibitem{badretriever}
Q.~Long, Y.~Deng, L.~Gan, W.~Wang, and S.~J. Pan, ``Backdoor attacks on dense passage retrievers for disseminating misinformation,'' 2024.

\bibitem{cheng2024trojanrag}
P.~Cheng, Y.~Ding, T.~Ju, Z.~Wu, W.~Du, P.~Yi, Z.~Zhang, and G.~Liu, ``Trojanrag: Retrieval-augmented generation can be backdoor driver in large language models,'' \emph{arXiv preprint arXiv:2405.13401}, 2024.

\bibitem{backdooricl1}
N.~Kandpal, M.~Jagielski, F.~Tramèr, and N.~Carlini, ``Backdoor attacks for in-context learning with language models,'' 2023.

\bibitem{backdooricl}
S.~Zhao, M.~Jia, L.~A. Tuan, F.~Pan, and J.~Wen, ``Universal vulnerabilities in large language models: Backdoor attacks for in-context learning,'' 2024.

\bibitem{foundationrisk}
R.~Bommasani, D.~A. Hudson, E.~Adeli \emph{et~al.}, ``On the opportunities and risks of foundation models,'' 2022.

\bibitem{anil2024many}
C.~Anil, E.~Durmus, M.~Sharma, J.~Benton, S.~Kundu, J.~Batson, N.~Rimsky, M.~Tong, J.~Mu, D.~Ford \emph{et~al.}, ``Many-shot jailbreaking,'' \emph{Anthropic, April}, 2024.

\bibitem{li2024badedit}
Y.~Li, T.~Li, K.~Chen, J.~Zhang, S.~Liu, W.~Wang, T.~Zhang, and Y.~Liu, ``Badedit: Backdooring large language models by model editing,'' \emph{arXiv preprint arXiv:2403.13355}, 2024.

\bibitem{backdoorsteering}
H.~Wang and K.~Shu, ``Trojan activation attack: Red-teaming large language models using activation steering for safety-alignment,'' 2024.

\bibitem{liu2017neural}
Y.~Liu, Y.~Xie, and A.~Srivastava, ``Neural trojans,'' in \emph{2017 IEEE International Conference on Computer Design (ICCD)}, 2017.

\bibitem{zeng2022adversarial}
Y.~Zeng, S.~Chen, W.~Park, Z.~Mao, M.~Jin, and R.~Jia, ``Adversarial unlearning of backdoors via implicit hypergradient,'' in \emph{ICLR}, 2022.

\bibitem{liu2018fine}
K.~Liu, B.~Dolan-Gavitt, and S.~Garg, ``Fine-pruning: Defending against backdooring attacks on deep neural networks,'' in \emph{21st International Symposium on Research in Attacks, Intrusions and Defenses, RAID 2018}, 2018.

\bibitem{liu-etal-2024-shortcuts}
Q.~Liu, F.~Wang, C.~Xiao, and M.~Chen, ``From shortcuts to triggers: Backdoor defense with denoised {P}o{E},'' in \emph{NAACL}, 2024, pp. 483--496.

\bibitem{graf2024two}
V.~Graf, Q.~Liu, and M.~Chen, ``Two heads are better than one: Nested poe for robust defense against multi-backdoors,'' in \emph{NAACL}, 2024.

\bibitem{wang2023robust}
\BIBentryALTinterwordspacing
F.~Wang, J.~Y. Huang, T.~Yan, W.~Zhou, and M.~Chen, ``Robust natural language understanding with residual attention debiasing,'' in \emph{Findings of the Association for Computational Linguistics: ACL 2023}, A.~Rogers, J.~Boyd-Graber, and N.~Okazaki, Eds.\hskip 1em plus 0.5em minus 0.4em\relax Toronto, Canada: Association for Computational Linguistics, Jul. 2023, pp. 504--519. [Online]. Available: \url{https://aclanthology.org/2023.findings-acl.32}
\BIBentrySTDinterwordspacing

\bibitem{li2021backdoor}
Y.~Li, T.~Zhai, Y.~Jiang, Z.~Li, and S.-T. Xia, ``Backdoor attack in the physical world,'' in \emph{ICLR Workshop}, 2021.

\bibitem{zhang2022fine}
Z.~Zhang, L.~Lyu, X.~Ma, C.~Wang, and X.~Sun, ``Fine-mixing: Mitigating backdoors in fine-tuned language models,'' in \emph{Findings: EMNLP}, 2022.

\bibitem{arora-etal-2024-heres}
A.~Arora, X.~He, M.~Mozes, S.~Swain, M.~Dras, and Q.~Xu, ``Here{'}s a free lunch: Sanitizing backdoored models with model merge,'' in \emph{Findings: ACL}, L.-W. Ku, A.~Martins, and V.~Srikumar, Eds., 2024.

\bibitem{qi-etal-2021-onion}
F.~Qi, Y.~Chen, M.~Li, Y.~Yao \emph{et~al.}, ``{ONION}: A simple and effective defense against textual backdoor attacks,'' in \emph{EMNLP}, 2021.

\bibitem{gao2019strip}
Y.~Gao, C.~Xu, D.~Wang, S.~Chen, D.~C. Ranasinghe, and S.~Nepal, ``Strip: A defence against trojan attacks on deep neural networks,'' in \emph{Proceedings of the 35th Annual Computer Security Applications Conference}, 2019.

\bibitem{subedar2019deep}
M.~Subedar, N.~Ahuja, R.~Krishnan, I.~J. Ndiour, and O.~Tickoo, ``Deep probabilistic models to detect data poisoning attacks,'' in \emph{NeurIPS Workshop}, 2019.

\bibitem{Du2020Robust}
M.~Du, R.~Jia, and D.~Song, ``Robust anomaly detection and backdoor attack detection via differential privacy,'' in \emph{ICLR}, 2020.

\bibitem{jin2020unified}
K.~Jin, T.~Zhang, C.~Shen, Y.~Chen, M.~Fan, C.~Lin, and T.~Liu, ``A unified framework for analyzing and detecting malicious examples of dnn models,'' \emph{arXiv preprint arXiv:2006.14871}, vol.~8, no.~9, 2020.

\bibitem{javaheripi2020cleann}
M.~Javaheripi, M.~Samragh, G.~Fields, T.~Javidi, and F.~Koushanfar, ``Cleann: Accelerated trojan shield for embedded neural networks,'' in \emph{ICCAD}, 2020.

\bibitem{mo2023test}
W.~Mo, J.~Xu, Q.~Liu, J.~Wang, J.~Yan, C.~Xiao, and M.~Chen, ``Test-time backdoor mitigation for black-box large language models with defensive demonstrations,'' \emph{arXiv preprint arXiv:2311.09763}, 2023.

\bibitem{li-etal-2023-defending}
J.~Li, Z.~Wu, W.~Ping, C.~Xiao, and V.~Vydiswaran, ``Defending against insertion-based textual backdoor attacks via attribution,'' in \emph{Findings: ACL}, A.~Rogers, J.~Boyd-Graber, and N.~Okazaki, Eds., 2023.

\bibitem{gao2021design}
Y.~Gao, Y.~Kim, B.~G. Doan, Z.~Zhang, G.~Zhang, S.~Nepal, D.~C. Ranasinghe, and H.~Kim, ``Design and evaluation of a multi-domain trojan detection method on deep neural networks,'' \emph{IEEE Transactions on Dependable and Secure Computing}, vol.~19, no.~4, 2021.

\bibitem{yang2021rap}
W.~Yang, Y.~Lin, P.~Li, J.~Zhou, and X.~Sun, ``Rap: Robustness-aware perturbations for defending against backdoor attacks on nlp models,'' \emph{arXiv preprint arXiv:2110.07831}, 2021.

\bibitem{chen2021mitigating}
C.~Chen and J.~Dai, ``Mitigating backdoor attacks in lstm-based text classification systems by backdoor keyword identification,'' \emph{Neurocomputing}, vol. 452, 2021.

\bibitem{he-etal-2023-mitigating}
X.~He, Q.~Xu, J.~Wang, B.~Rubinstein, and T.~Cohn, ``Mitigating backdoor poisoning attacks through the lens of spurious correlation,'' in \emph{EMNLP}, H.~Bouamor, J.~Pino, and K.~Bali, Eds., 2023.

\bibitem{fields2021trojan}
G.~Fields, M.~Samragh, M.~Javaheripi, F.~Koushanfar, and T.~Javidi, ``Trojan signatures in dnn weights,'' in \emph{ICCV}, 2021.

\bibitem{lyu-etal-2022-study}
W.~Lyu, S.~Zheng, T.~Ma, and C.~Chen, ``A study of the attention abnormality in trojaned {BERT}s,'' in \emph{NAACL}, M.~Carpuat, M.-C. de~Marneffe, and I.~V. Meza~Ruiz, Eds., 2022.

\bibitem{xu2021detecting}
X.~Xu, Q.~Wang, H.~Li, N.~Borisov, C.~A. Gunter, and B.~Li, ``Detecting ai trojans using meta neural analysis,'' in \emph{S\&P}, 2021.

\bibitem{mazeika2022trojan}
M.~Mazeika, D.~Hendrycks, H.~Li, X.~Xu, S.~Hough, A.~Zou, A.~Rajabi, Q.~Yao, Z.~Wang, J.~Tian \emph{et~al.}, ``The trojan detection challenge,'' in \emph{NeurIPS 2022 Competition Track}, 2022.

\bibitem{azizi2021t}
A.~Azizi, I.~A. Tahmid, A.~Waheed, N.~Mangaokar, J.~Pu, M.~Javed, C.~K. Reddy, and B.~Viswanath, ``$\{$T-Miner$\}$: A generative approach to defend against trojan attacks on $\{$DNN-based$\}$ text classification,'' in \emph{USENIX Security}, 2021.

\bibitem{liu2019abs}
Y.~Liu, W.-C. Lee, G.~Tao, S.~Ma, Y.~Aafer, and X.~Zhang, ``Abs: Scanning neural networks for back-doors by artificial brain stimulation,'' in \emph{CCS}, 2019.

\bibitem{whitehouse2023factsheet}
\BIBentryALTinterwordspacing
``Fact sheet: President biden issues executive order on safe, secure, and trustworthy artificial intelligence,'' 2023. [Online]. Available: \url{https://www.whitehouse.gov/briefing-room/statements-releases/2023/10/30/fact-sheet-president-biden-issues-executive-order-on-safe-secure-and-\trustworthy-artificial-intelligence/}
\BIBentrySTDinterwordspacing

\bibitem{pelc2024cybersecurity}
P.~Pelc, ``Cybersecurity issue in the executive order on the safe, secure, and trustworthy development and use of artificial intelligence from october 30, 2023,'' \emph{Cybersecurity and Law}, vol.~11, no.~1, 2024.

\bibitem{chen2024multi}
B.~Chen, N.~Ivanov, G.~Wang, and Q.~Yan, ``Multi-turn hidden backdoor in large language model-powered chatbot models,'' in \emph{CCS}, 2024.

\bibitem{hao2024exploring}
Y.~Hao, W.~Yang, and Y.~Lin, ``Exploring backdoor vulnerabilities of chat models,'' \emph{arXiv preprint arXiv:2404.02406}, 2024.

\bibitem{kandpalbackdoor}
N.~Kandpal, M.~Jagielski, F.~Tram{\`e}r, and N.~Carlini, ``Backdoor attacks for in-context learning with language models,'' in \emph{The Second Workshop on New Frontiers in Adversarial Machine Learning}.

\bibitem{zhao2024universal}
S.~Zhao, M.~Jia, L.~A. Tuan, F.~Pan, and J.~Wen, ``Universal vulnerabilities in large language models: Backdoor attacks for in-context learning,'' \emph{arXiv preprint arXiv:2401.05949}, 2024.

\bibitem{xu2024cognitive}
N.~Xu, F.~Wang, B.~Zhou, B.~Li, C.~Xiao, and M.~Chen, ``Cognitive overload: Jailbreaking large language models with overloaded logical thinking,'' in \emph{Findings: NAACL}, 2024.

\bibitem{carlini2024poisoning}
N.~Carlini, M.~Jagielski, C.~A. Choquette-Choo, D.~Paleka, W.~Pearce, H.~Anderson, A.~Terzis, K.~Thomas, and F.~Tram{\`e}r, ``Poisoning web-scale training datasets is practical,'' in \emph{IEEE S\&P}, 2024.

\bibitem{bowen2024scaling}
D.~Bowen, B.~Murphy, W.~Cai, D.~Khachaturov, A.~Gleave, and K.~Pelrine, ``Scaling laws for data poisoning in llms,'' \emph{arXiv preprint arXiv:2408.02946}, 2024.

\bibitem{wang2024data}
F.~Wang, N.~Mehrabi, P.~Goyal, R.~Gupta, K.-W. Chang, and A.~Galstyan, ``Data advisor: Constitutional data curation for safety alignment of large language models,'' in \emph{EMNLP}, 2024.

\bibitem{wang2023causal}
F.~Wang, W.~Mo, Y.~Wang, W.~Zhou, and M.~Chen, ``A causal view of entity bias in (large) language models,'' \emph{EMNLP - Findings}, 2023.

\bibitem{dong2021black}
Y.~Dong, X.~Yang, Z.~Deng, T.~Pang, Z.~Xiao, H.~Su, and J.~Zhu, ``Black-box detection of backdoor attacks with limited information and data,'' in \emph{ICCV}, 2021.

\bibitem{li2023multi}
Y.~Li, S.~Liu, K.~Chen, X.~Xie, T.~Zhang, and Y.~Liu, ``Multi-target backdoor attacks for code pre-trained models,'' in \emph{ACL}, 2023.

\bibitem{yuan2024rigorllm}
Z.~Yuan, Z.~Xiong, Y.~Zeng, N.~Yu, R.~Jia, D.~Song, and B.~Li, ``Rigorllm: Resilient guardrails for large language models against undesired content,'' \emph{arXiv preprint arXiv:2403.13031}, 2024.

\bibitem{yang2024stealthy}
Z.~Yang, B.~Xu, J.~M. Zhang, H.~J. Kang, J.~Shi, J.~He, and D.~Lo, ``Stealthy backdoor attack for code models,'' \emph{TSE}, 2024.

\bibitem{wu2023deceptprompt}
F.~Wu, X.~Liu, and C.~Xiao, ``Deceptprompt: Exploiting llm-driven code generation via adversarial natural language instructions,'' \emph{arXiv preprint arXiv:2312.04730}, 2023.

\bibitem{liu2024rethinking}
T.~Liu, F.~Wang, and M.~Chen, ``Rethinking tabular data understanding with large language models,'' in \emph{NAACL}, 2024, pp. 450--482.

\bibitem{wang2022robust}
F.~Wang, Z.~Xu, P.~Szekely, and M.~Chen, ``Robust (controlled) table-to-text generation with structure-aware equivariance learning,'' in \emph{NAACL}, 2022.

\end{thebibliography}

\end{document}